\documentclass{cimento}
\usepackage{amsmath}
\usepackage{amssymb}
\usepackage{graphicx}
\usepackage{bm}
\usepackage{multirow}
\usepackage{xspace}
\usepackage{color}

\newcommand{\jpsi}{\ensuremath{J/\psi}\xspace}

\newcommand{\swave}{\ensuremath{S\text{-wave}}\xspace}
\newcommand{\pwave}{\ensuremath{P\text{-wave}}\xspace}
\newcommand{\helium}{\ensuremath{^3\text{He}}\xspace}

\newcommand{\Zcprime}{\ensuremath{Z_c^\prime}\xspace}

\newcommand{\babar}{\mbox{\slshape B\kern-0.1em{\smaller A}\kern-0.1em
    B\kern-0.1em{\smaller A\kern-0.2em R}}\xspace}

\title{$XYZ$: four quark states?}
\author{A.~Pilloni}
\instlist{\inst{} Dipartimento di Fisica and INFN, `Sapienza' Universit\`a di Roma\\
P.le Aldo Moro 2 - I-00185 Roma (Italy)
  }

\PACSes{\PACSit{14.40.Rt}{Exotic mesons}
\PACSit{12.39.Jh}{Nonrelativistic quark model}
\PACSit{25.75.Dw}{Particle and resonance production}}

\begin{document}

\maketitle

\begin{abstract}
The observation of many unexpected states decaying into heavy quarkonia has challenged the usual $Q \bar Q$ interpretation. We will discuss the nature of some of the charmonium-like resonances recently observed by BES~III and LHCb, and their identification according to the compact tetraquark model. 
We also comment the production of light nuclei in hadron collisions and the relevance for the physics of the $X(3872)$.
\end{abstract}

\begin{figure}[t]
\centering
\includegraphics[width=.6\textwidth]{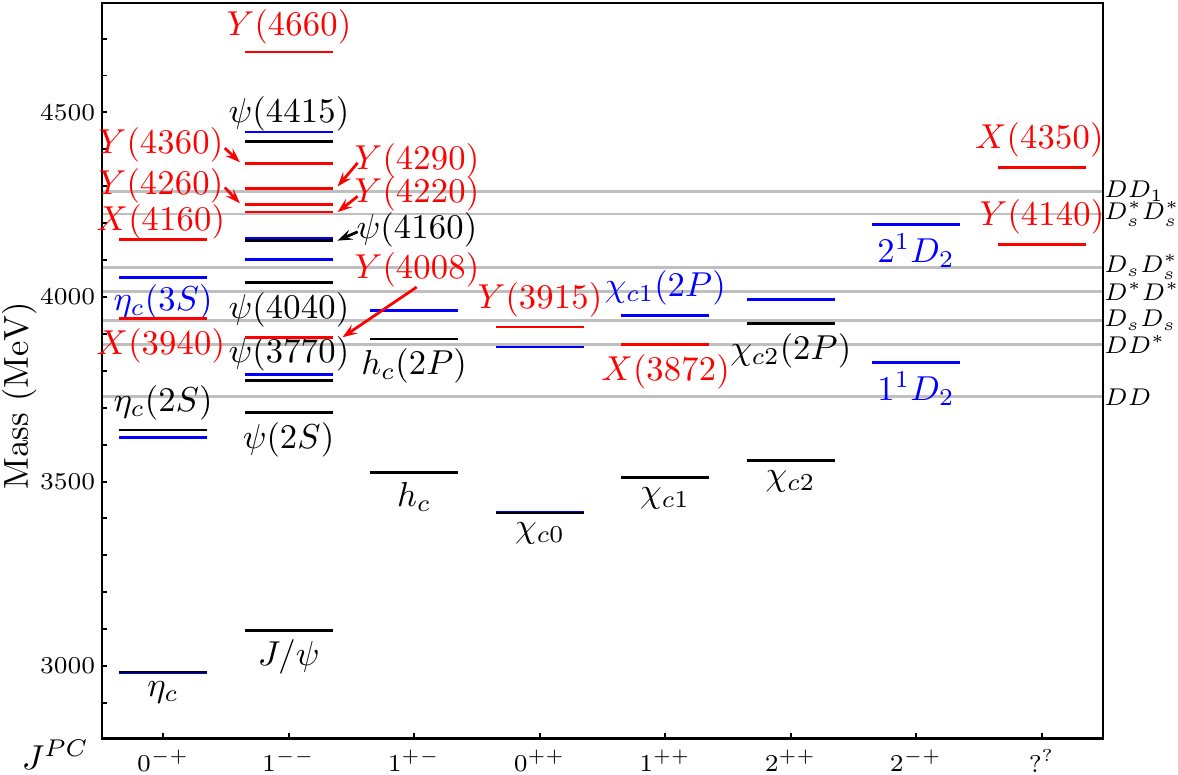}
\caption{Charmonium sector, from~\cite{Esposito:2014rxa}.
The black lines represent observed charmonium levels, the blue lines represent 
predicted levels, and
the red ones are exotic states. The open charm thresholds are reported on the right.}.
 \label{fig:charmonium_levels}
\end{figure}

{\bf \em Introduction} ---
In the last ten years lots of unexpected $XYZ$ resonances have been discovered in the heavy quarkonium sector.  
Their production and decay rates are not compatible with a standard quarkonium interpretation. The resulting charmonium spectrum is summarized in Fig.~\ref{fig:charmonium_levels}. 
Among the most likely phenomenological interpretation, we recall:
$i)$~{\em molecule:} bound state of two mesons, interacting via long-range light meson exchange; 
$ii)$~{\em tetraquark:} compact state made of a diquark (a $qq$ bound pair in the $\bar {\bm 3}_c$) and of an antidiquark; 
$iii)$~{\em hybrid:} state of quarks and constituent gluons; 
$iv)$~{\em hadroquarkonium:} heavy $Q\bar Q$ pair surrounded by light hadronic matter.
Here we will focus on the compact tetraquark model, along the lines presented in~\cite{Maiani:2004vq,Maiani:2014aja}. 
For a review, see~\cite{Esposito:2014rxa,Faccini:2012pj}.

\begin{table}[t]
\begin{tabular}{lllll}
    \hline
$J^{PC}$ & $cq\;\bar c\bar q$&$c\bar c\;q\bar q$&Resonance Assig. &Decays\\ \hline
$0^{++}$ & $|0,0\rangle$ & $1/2|0,0\rangle+ \sqrt{3}/2|1,1\rangle_0$ & $X_0 (\sim 3770~\mathrm{MeV})$ & $\eta_c, \jpsi$ + light had.\\
$0^{++}$ & $|1,1\rangle_0$ &  $\sqrt{3}/2|0,0\rangle-1/2|1,1\rangle_0$ & $X_0^\prime (\sim 4000~\mathrm{MeV})$ & $\eta_c,\jpsi$ + light had.\\
$1^{++}$ & $(|1,0\rangle+|0,1\rangle)/\sqrt{2}$ & $|1,1\rangle_1$ & $X(3872)$& $\jpsi+\rho/\omega$, $DD^*$\\
$1^{+-}$ & $(|1,0\rangle-|0,1\rangle)/\sqrt{2} $ &  $(|1,0\rangle-|0,1\rangle)/\sqrt{2} $ & $Z_c(3900)$ & $\jpsi \,\pi,\,h_c\,\pi, \eta_c\,\rho$\\
$1^{+-}$ & $|1,1\rangle_1$ & $(|1,0\rangle+|0,1\rangle)/\sqrt{2} $ & $Z_c^\prime(4020)$ & $\jpsi \,\pi,\,h_c\,\pi, \eta_c\,\rho$\\
$2^{++}$ & $|1,1\rangle_2$ & $|1,1\rangle_2$ & $X_2(\sim4000~\mathrm{MeV})$ & $\jpsi$ + light had.\\\hline
    \end{tabular}
\begin{tabular}{cccc}
    \hline
State &$P(S_{c\bar c}=1):P(S_{c\bar c}=0)$& Assignment & Radiative Decay\\ \hline
$Y_1$& 3:1 & $Y(4008)$ & $\gamma +X_0$ \\
$Y_2$&1:0 & $Y(4260)$ & $\gamma+X$\\
$Y_3$&1:3 & $Y(4290)/Y(4220)$ & $\gamma+ X_0^\prime$ \\
$Y_4$&1:0 & $Y(4630)$ & $\gamma+X_2$\\
\hline
    \end{tabular}
\caption{Summary of the $L=0$ and $L=1$ tetraquarks}\label{primatab}
\end{table}
{\bf \em Tetraquark and pentaquark candidates} --- In the region $3850$-$4050$~MeV, three axial exotic resonances have been observed. The most famous one is the $X(3872)$, discovered in the $B \to K (\jpsi\,\pi\pi)$ channel. The mass is very close to the $\bar D^0 D^{*0}$ threshold, with $\Delta M = -3 \pm 192$~keV~\cite{Tomaradze:2015cza}, and the width is much more narrow than the experimental resolution, $\Gamma < 1.2$~MeV at 90\% C.L.~\cite{Choi:2011fc}. After some controversies~(see \cite{Faccini:2012zv}), the quantum numbers are now well established to be $J^{PC} = 1^{++}$~\cite{Aaij:2013zoa}. In the molecular picture, the $\Delta M$ would be the binding energy, which has to be negative, and is related to the coupling to the constituent; testing this relation is still beyond the present experimental accuracy~\cite{Polosa:2015tra}. The $\pi\pi$ pair is dominated by the isovector $\rho$ resonance, which imply a large isospin violation. Other two charged states, with $J^{PC} = 1^{+-}$, have been seen at lepton colliders: the $Z_c(3900)^+$ has been found in the $Y(4260) \to (\jpsi \,\pi^+) \pi^-$ channel~\footnote{Hereafter the charge-conjugated modes are understood.}, with mass and width $M =  3888.7 \pm 3.4$~MeV,  $\Gamma = 35\pm7$~MeV; the $\Zcprime(4020)^+$ has been found in the $e^+e^- \to (h_c \,\pi^+) \pi^-$ channel, with mass and width $M = 4023.9 \pm 2.4$~MeV, $\Gamma = 10\pm 6$~MeV. Both state are close to the $(D D^*)^+$ and $\bar D^{*0} D^{*+}$ thresholds, respectively, and are indeed observed to decay into those open-charm pairs.
The interpretation in terms of tetraquarks~\cite{Maiani:2004vq,Maiani:2014aja,Faccini:2013lda} accomodates many properties of these states. In particular, it would explain the isospin violation in the $X(3872)$ decays, by means of a mechanism proposed many years ago for baryonia states~\cite{Rossi:1977dp,Montanet:1980te}. The confirmation of the $Z(4430)^+$ in the $\bar B^0 \to K^- (\psi(2S)\,\pi^+)$ channel~\cite{Aaij:2014jqa}, and its identification as the radial excitation of the $Z_c(3900)$~\cite{Maiani:2014aja,Maiani:2007wz} gives more strength to the tetraquark framework. In~\cite{Esposito:2014hsa}, it has been proposed to seek the $Z_c$ and the $Z_c^\prime$ in the $\eta_c\,\rho$ final states, which should be favored according to the tetraquark hypothesis, and suppressed in the molecular picture. 

Lepton colliders have also reported the observation of some $J^{PC}=1^{--}$ states produced in association with an ISR photon. The most famous is the $Y(4260)$, seen as a resonance in the $\jpsi \,\pi\pi$ invariant mass. The missing observation of the decay $Y(4260) \to D \bar D$ prevents the identification as an ordinary charmonium. A similar structure has been observed in the $h_c \,\pi^+\pi^-$~\cite{Ablikim:2013wzq} and $\chi_{c0}\,\omega$~\cite{Ablikim:2014qwy} invariant masses, but with a lineshape not compatible with the $Y(4260)$ one. This new state has been named $Y(4220)$~\cite{yuan1}, and the apparent heavy quark spin symmetry violation might be accomodated in the tetraquark picture~\cite{Faccini:2014pma}. The $Y(4008)$, seen in $\jpsi\,\pi^+\pi^-$, and the $Y(4630)$, seen in $\Lambda_c^+ \Lambda_c^-$, complete this new tetraquark multiplet, which can be naturally interpreted as the $L=1$ orbital excitation of the multiplet we discussed before~\cite{Maiani:2014aja}. The observation of $Y(4260) \to \gamma\, X(3872)$~\cite{Ablikim:2013dyn} favors this picture, being the natural $E1$ electromagnetic emission. The identifications are summarized in Tab.~\ref{primatab}.
Moreover, in this picture the two $Y(4360)$ and $Y(4660)$ resonances seen in the $\psi(2S)\,\pi^+\pi^-$ invariant mass can be identified as the radial excitations of the $Y(4008)$ and $Y(4260)$, respectively.

Exotic states have also been sought on the lattice. Some work has been performed in the last two years, and found evidence for an $X(3872)$ state~\cite{Prelovsek:2013cra}, and no signal of any other states, including the charged ones~\cite{Prelovsek:2013xba,Prelovsek:2014swa,Padmanath:2015era}. These difficult studies are still affected by large systematics and finite volume effects, and are far from being conclusive on the nature of the exotic states. A preliminary study of doubly charmed states has been also started~\cite{Esposito:2013fma,Guerrieri:2014nxa}.  

Finally, we recall the recent observation of two pentaquark candidates by LHCb, in the decay $\Lambda_b^0 \to \jpsi\,p\,K^-$ as a resonance in the $\jpsi \,p$ channel~\cite{Aaij:2015tga}. The lighter state has mass and width $M_1 =  4380 \pm 8 \pm 29$~MeV and $\Gamma_1 = 205 \pm 18 \pm 86$~MeV, and most likely signature $J^P = \frac32^-$, the heavier one has $M_2 = 4449.8 \pm 1.7 \pm 2.5$~MeV, $\Gamma_2 = 39 \pm 5 \pm 19$~MeV, and $J^P = \frac52^+$. Two states so close in mass but with different parities can hardly be explained by molecular models, because of the lack of open-charm thresholds with the correct quantum numbers. An interpretation in terms of a diquark-diquark-antiquark system has been proposed in~\cite{Maiani:2015vwa}: the lighter state is an \swave pentaquark, the heavier is a \pwave one. The mass difference is expected to be $\sim 300$~MeV, but can be reduced to $\sim 100$~MeV if the two states have diquark content
\begin{align*}
 P_1 &= \Big(\bar c \,[c u]_{S=1} [d u]_{S=1}  \Big)_{L=0},   & P_2 &=  \Big(\bar c\,[c u]_{S=1} [d u]_{S=0} \Big)_{L=1}
\end{align*}

A deeper analysis with Run~II statistics is needed to better establish the Breit-Wigner parameters of these states, and to look for other possible  broad peaks in the same region.

\begin{figure}[t]
 \begin{center}
   \includegraphics[width=.5\textwidth]{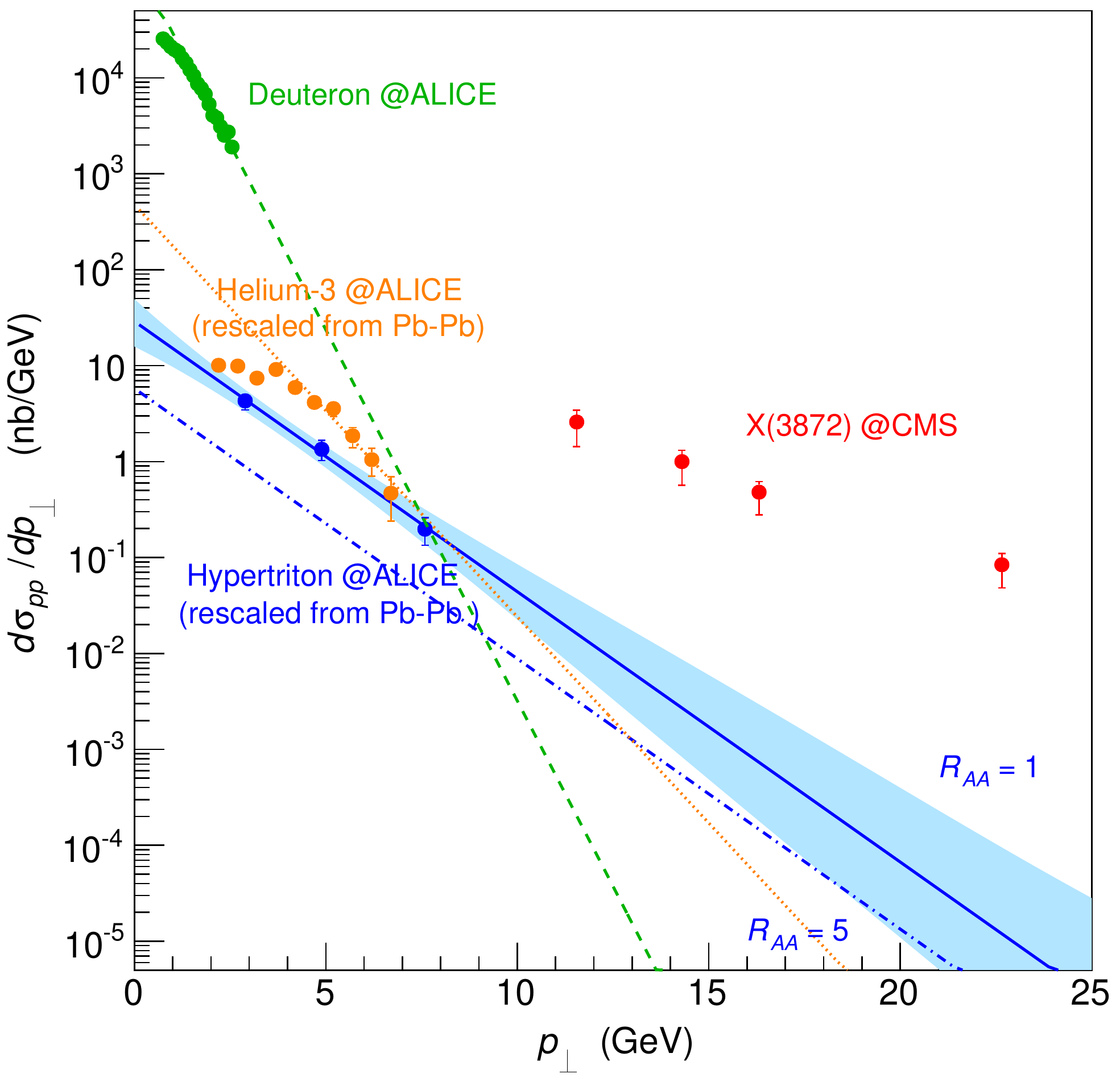}
 \end{center}
\caption{\small Comparison between the prompt production cross section in $pp$ collisions of $X(3872)$ (red), deuteron (green), \helium (orange), and hypertriton (blue), from~\cite{Esposito:2015fsa}. The $X$ data from CMS~\cite{Chatrchyan:2013cld} are rescaled by the branching ratio $\mathcal{B}(X\to J/\psi\,\pi\pi)$. Deuteron data in $pp$ collisions are taken from ALICE~\cite{Adam:2015vda}.
The \helium and hypertriton data measured by ALICE in Pb-Pb collisions~\cite{Adam:2015yta,Adam:2015vda} have been rescaled to $pp$ using a Glauber model. 
The dashed green line is the exponential fit to the deuteron data points in the $p_\perp\in[1.7,3.0]$ GeV region, whereas the dotted orange one is the fit to the \helium data points.  The solid and dot-dashed blue lines represent the fits to hypertriton data with $R_{AA}=1$ (no medium effects) and an hypothetical constant value of $R_{AA}=5$. \label{uno}
 The hypertriton data are fitted with an exponential curve, and the light blue band is the $68\%$~C.L. for the extrapolated $R_{AA}=1$ curve. \helium data in the $p_\perp \in[4.45,6.95]$ GeV region are also fitted with an exponential curve. 
}
\end{figure}
{\bf \em Comparison between $X(3872)$ and light nuclei at hadron colliders} ---
The copious prompt production of $X(3872)$ at hadron colliders is the main drawback of any molecular interpretation: how is that possible that a molecule of a $\bar D^0$ and a $D^{*0}$ meson, with binding energy compatible with zero, could be formed within the hadrons ejected in $pp$ collisions at energies of some TeV?  Indeed, CMS reported a production cross section of $\sim 13$~nb, at $p_\perp > 15$~GeV~\cite{Chatrchyan:2013cld}. A simple estimate with usual MC generators gives an upper bound for the cross sections two orders of magnitude smaller than the experimental value~\cite{Bignamini:2009sk}. To bypass this result, people had recourse to Final State Interactions~\cite{Artoisenet:2009wk}, but the application in high energy collisions led to some controversies~\cite{Bignamini:2009fn,Artoisenet:2010uu}. An alternative mechanism to increase the cross section was explored in~\cite{Esposito:2013ada,Guerrieri:2014gfa}, but still not enough to reach the experimental value.

Moreover, in~\cite{Guerrieri:2014gfa} it was proposed to compare the production of deuteron, a {\em bona fide} hadron molecule, with the $X(3872)$ one: if the $X(3872)$ were a deuteron-like molecule, a similar production cross section is expected, regardless of the details of any mechanism needed to increase the MC results.

Very recently, ALICE reported the observation of light nuclei in $pp$ and Pb-Pb collisions~\cite{Adam:2015yta,Adam:2015vda}. Although a proper comparison would require the measurement of light nuclei production in $pp$ collisions only, and at the same $p_\perp > 15$~GeV where the $X$ is seen, a simple exponential extrapolation of available data has been performed~\cite{Esposito:2015fsa}. Data in Pb-Pb collisions have been extrapolated to $pp$ by means of a Glauber MC, and of a naive rescaling from $\sqrt{s}= 2.76$~TeV to $\sqrt{s} = 7$~TeV. The results are reported in Fig.~\ref{uno}: in particular we appreciate that the hypertriton cross section is $2\div 3$ orders of magnitude smaller than the $X(3872)$ one, challenging a similar identification for the two states. The proper inclusion of medium effects, neglected in the Glauber approach, would even increase this gap. We stress that for an unbiased and definitive comparison with $X$ production, light nuclei should be searched in $pp$ collisions rather than in Pb-Pb, and at $p_\perp$ as high as 15~GeV. These analyses can be performed by ALICE and LHCb during Run~II.

\acknowledgments
I wish to thank the organizers and in particular N.~Tantalo for their kind invitation. I also thank G.C.~Rossi for useful discussions on baryonia.

\bibliographystyle{varenna}

\end{document}